\documentclass[a4paper,gaps,aps,prb,10pt,showpacs,twocolumn,superscriptaddress,showkeys]{revtex4-2}
\usepackage{setspace}
\usepackage{amsmath,amssymb,mathrsfs}
\usepackage[utf8]{inputenc}
\usepackage[pdftex]{graphicx}
\usepackage[dvipsnames]{xcolor}
\usepackage[normalem]{ulem}
\usepackage[british]{babel}
\usepackage{csquotes}
\usepackage{comment}
\usepackage[version=4]{mhchem}
\usepackage{silence}
\usepackage[colorlinks,citecolor=red,urlcolor=blue,bookmarks=false,hypertexnames=true]{hyperref}

\DeclareUnicodeCharacter{2212}{-}

\newcommand{\QUB}{School of Mathematics and Physics, Queen's University Belfast, Belfast BT7 1NN, Northern Ireland, United Kingdom}
\newcommand{\ETSF}{European Theoretical Spectroscopy Facility (ETSF)}
\newcommand{\IPPT}{Institute of Fundamental Technological Research, Polish Academy of Sciences, Adolfa Pawińskiego 5b, 02-106 Warsaw, Poland}
\begin{document}

\title{Interplay of spin-phonon coupling in FeSe with pressure and isoelectronic (S/Te) substitution}

\author{Abyay Ghosh}
\affiliation{\QUB}
\author{Piotr Chudzinski} 
\affiliation{\QUB}
\affiliation{\IPPT}
\author{Myrta Gr\"uning}
\affiliation{\QUB}
\affiliation{\ETSF}
\date{\today}

\date{\today}

\begin{abstract}

In several studies, magnetism has been shown to strongly influence the properties of the iron chalcogenide compound \ce{FeSe}. In our previous work [Phys. Rev. Res. 6(4), 043154 (2024)], we demonstrated how the interplay between spin and orbital degrees of freedom governs the electronic structure and consequently shapes the magnetic phase diagram of \ce{FeSe} under external hydrostatic pressure. However, a comprehensive understanding of the phase diagram also requires investigating the role of lattice degrees of freedom (phonons) and their coupling to magnetism. In the present study, we examine how hydrostatic pressure and isoelectronic substitution affect the magnetic and lattice dynamical properties of \ce{FeSe} using Density Functional Theory calculations. Our lattice dynamics analysis reveals that different phonon modes exhibit distinct dependencies on the local magnetic moment across various magnetic phases. We further identify pressure- and magnetic-moment-dependent dynamical instabilities in specific magnetic configurations, highlighting the intricate coupling between structural and magnetic degrees of freedom in \ce{FeSe}. An investigation of spin-phonon coupling demonstrates the presence of strong magnetoelastic interactions affecting the prominent optical phonon modes: the Fe-derived $B_{1g}$ mode and the Se-derived $A_{1g}$ mode. Finally, we discuss the evolution of spin-phonon coupling under pressure and alloying in the context of its possible implications for emergent phases such as magnetism, nematicity, and superconductivity in \ce{FeSe}.
\end{abstract}

\keywords{First principles calculation; iron based superconductors; magnetism; lattice dynamics; phonon modes; spin-phonon coupling}
\maketitle

\section{Introduction}

Materials with a layered quasi-two-dimensional structure often allow the presence of various strong energy coupling terms giving rise to macroscopic interplay between spin, orbital, and lattice degrees of freedom \cite{a,PhysRevMaterials.7.045201}. This in turn can be manifested as magneto-elastic coupling in the material and is central to the physics of iron-based superconductors \cite{PhysRevResearch.2.033140,Marques2025}. Among them, \ce{FeSe} stands out for exhibiting a structural (nematic) transition without long-range magnetic order, yet it remains far from being magnetically inert. Theoretical studies propose competing hidden magnetic orders at ambient pressure~\cite{Glasbrenner2015,cite-key6,Fernandes2022,PhysRevB.94.035108}, and neutron spin resonance experiments reveal both N\'eel- and stripe-type fluctuations~\cite{Wang2016}, underscoring the presence of dynamic magnetism \cite{Kothapalli2016}. The strong dependence of magnetic exchange energies on the local Fe moment~\cite{PhysRevResearch.6.043154,PhysRevLett.124.117001} suggests a delicate feedback between magnetism, lattice, and electronic structure. {\em Understanding how these magnetic instabilities evolve under pressure or alloying, and how they feed back onto the structural degrees of freedom, is therefore key to unraveling the magnetoelastic interactions in \ce{FeSe}}. While density functional theory (DFT) has been instrumental in modeling iron-based superconductors~\cite{PhysRevB.78.134514,cite-key7a,Yamada_2022,Liu_2015}, systematic DFT investigations capturing the pressure- and alloying-dependence of phonons remain scarce in \ce{FeSe}. Existing approaches present a trade-off: field-theory models like in Ref.~\cite{PhysRevResearch.6.043154} neglect lattice dynamics, while studies like in Ref.~\cite{10.1039/d1cp02749b} analyze electron-phonon coupling purely from structural perspective without accounting for magnetism. Neither framework resolves how magnetic phases maintain dynamic stability or feed back into phonon behavior under pressure and alloying.


 
 Both questions are addressed in this work. First, we inspect whether magnetic orders survive dynamic coupling with the lattice. Second, we analyse the coupling between spin-orbital and phonon fluctuations. We study how both the dynamic stability and spin-phonon coupling evolve with external hydrostatic pressure and isoelectronic substitution. Overall, this study provides a comprehensive framework for understanding the interplay among orbital, spin, and lattice degrees of freedom, helping to rationalize the phase diagram of FeSe.

In Sec.~\ref{sec:method}, we present the computational methodology of the calculations performed. In Sec.~\ref{sec:phonon}, we study the phonon dispersion of \ce{FeSe} with pressure (Sec.\ref{subsec:p}), magnetic moment (Sec.\ref{subsec:m}) and alloying (Sec.\ref{subsec:s})in different magnetic phases. We assess the stability of different magnetic phases at higher pressures consistently with the FeSe phase diagram.  We observe a non-trivial dependence of phonon dispersion on the Fe magnetic moment with different spin configuration which hints toward strong spin-phonon coupling in this compound.
We thus provide a detailed analysis of the nature of spin-phonon coupling for the phonon modes $B_{1g}$ and $A_{1g}$ (Sec.~\ref{sec:spc}) and discuss its implications on the phase diagram (Sec.\ref{sec:spcdis}).

\section{Computational Details}\label{sec:method}
The plane wave pseudopotential suite QUANTUM ESPRESSO~\cite{Giannozzi_2009,Giannozzi_2017} is used to perform fully self-consistent DFT-based electronic structure calculations by solving the standard Kohn-Sham (KS) equations. Ultrasoft pseudopotentials from the PSlibrary~\cite{DALCORSO2014337} are used for Fe, Se, S and Te atoms. Kinetic-energy cut-offs are fixed to 55 Ry for electronic wave functions after performing rigorous convergence tests.

The electronic exchange-correlation is treated under the generalized gradient approximation (GGA) that is parametrized by Perdew-Burke-Enzerhof (PBE) functional~\cite{PhysRevLett.77.3865,PhysRevLett.78.1396}. 
The phonon dispersion is calculated with the finite displacement method using PHONOPY~\cite{phonopy-phono3py-JPCM,phonopy-phono3py-JPSJ} as well as Density Functional Perturbation Theory (DFPT)~\cite{PhysRevLett.58.1861} as implemented in QUANTUM ESPRESSO. A dense $q$-mesh grid of 3$\times$3$\times$3 is considered for the DFPT calculation. 
 
The model considered to simulate different magnetic phases like ferromagnetic (FM), checkerboard antiferromagnetic (CAFM), striped antiferromagnetic (SAFM) and staggered dimer (SD) has been discussed in \cite{supp1} in detail. We adopt the Monkhorst-Pack scheme~\cite{PhysRevB.13.5188} to sample the Brillouin zone in k-space with 8$\times$8$\times$8 grid. Geometry optimization has been performed using the Broyden-Fletcher-Goldfrab-Shanno (BFGS) scheme~\cite{BFGS_Opt}. The experimental lattice parameters ($a,b$ = 3.7698 $\AA$, $c =$$ 5.5163 \AA$ and $z_{\ce{Se}}$ = 0.2576) are used as starting values. The Convergence threshold of $10^{-8}$ and $10^{-5}$ are used on total energy (a.u) and forces (a.u) respectively for ionic minimization. High-pressure structures are obtained by enthalpy ($H=U+PV$) minimization under externally applied hydrostatic pressure. Energy penalty functional is used to perform the constrained magnetic moment calculations. The penalty term is incorporated into total energy by weight $\lambda$ as: $E_\text{total}=E_\text{LSDA}+\sum_{i}\lambda(M_{i}-M^0_{i})^2$, where $i$ is the atomic index for Fe atoms and $M^0_{i}$, $M_{i}$ are the targeted and actual local magnetic moment at atom $i$ respectively. The value of $\lambda$ is fixed to 25$Ry/\mu^2_{B}$ after performing a convergence test, constraining the magnetic moment of Fe at a particular value. The relevant details are included in \cite{supp1}. For the lattice dynamics calculation with S/Te isovalent substitution, conventional unit cell of FeSe is used where one Se atom is substituted with S/Te.

\section{Lattice dynamics}\label{sec:phonon}

\begin{figure}
 \centering
 \includegraphics[width=0.48\textwidth]{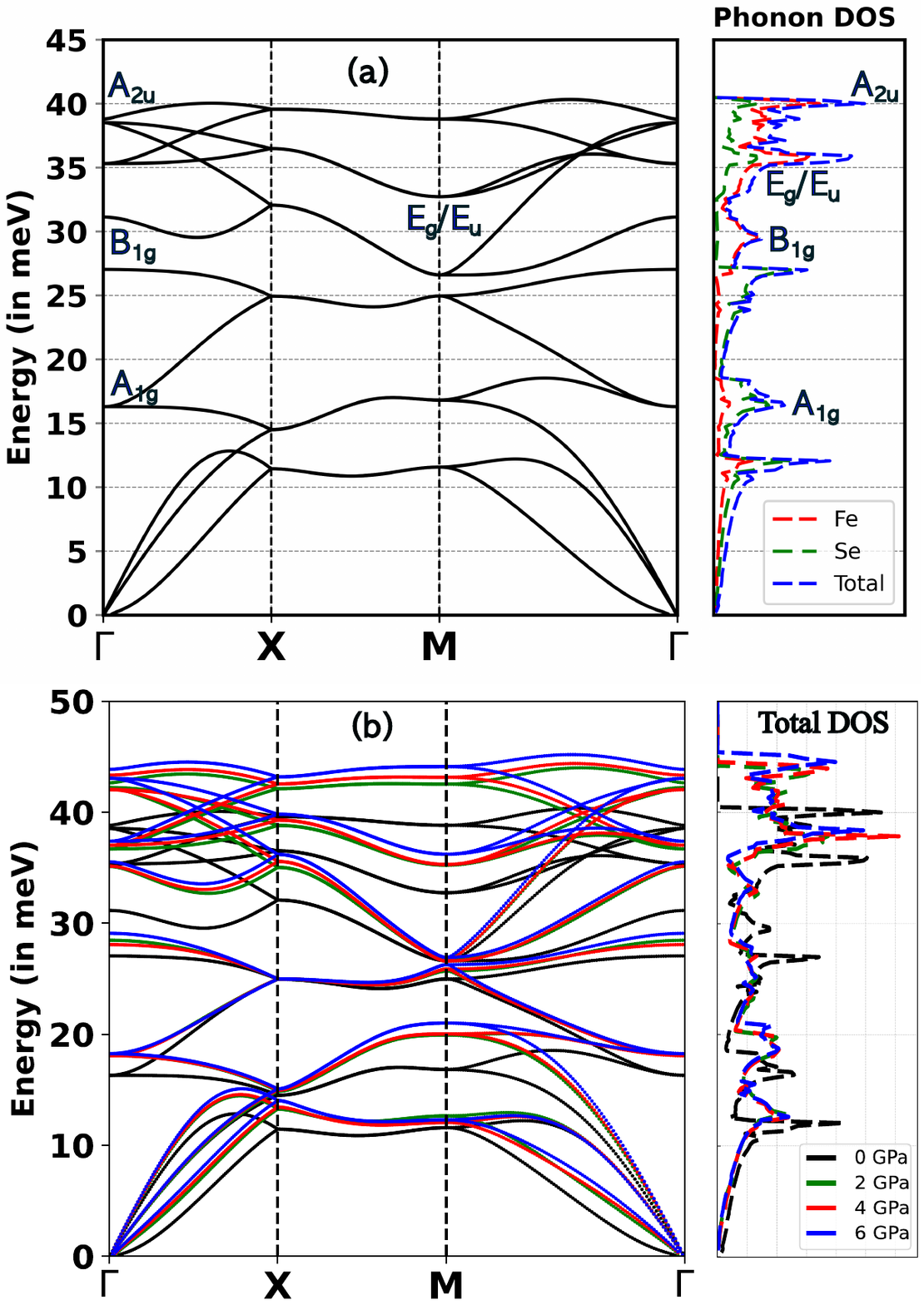}
 \caption{(a) Phonon dispersion and atom resolved phonon DOS marking different phonon modes in nonmagnetic phase at ambient pressure, (b) the change in phonon dispersion and DOS under hydrostatic pressure (0 GPa: \enquote{black}; 2 GPa: \enquote{green}; 4 GPa: \enquote{red}; 6 GPa: \enquote{blue}) in the nonmagnetic phase. }  
 \label{press}
\end{figure}

Figure~\ref{press} shows the calculated phonon dispersion and density of states (DOS) at ambient pressure of the nonmagnetic phase. The phonon modes are identified and marked in Fig.~\ref{press}(a) by looking at the atom-resolved phonon DOS. We identified the phonon modes which are observed in high-resolution electron energy-loss spectroscopy (HREELS) experiment~\cite{PhysRevB.96.094531}. There are three identified modes, $A_{1g}$,  $B_{1g}$ and  $A_{2u}$, around the $\Gamma$-point and two modes, $E_g$, $E_u$ around Brillouin zone corner M-point.
\begin{figure}
\centering
\includegraphics[width=0.40\textwidth]{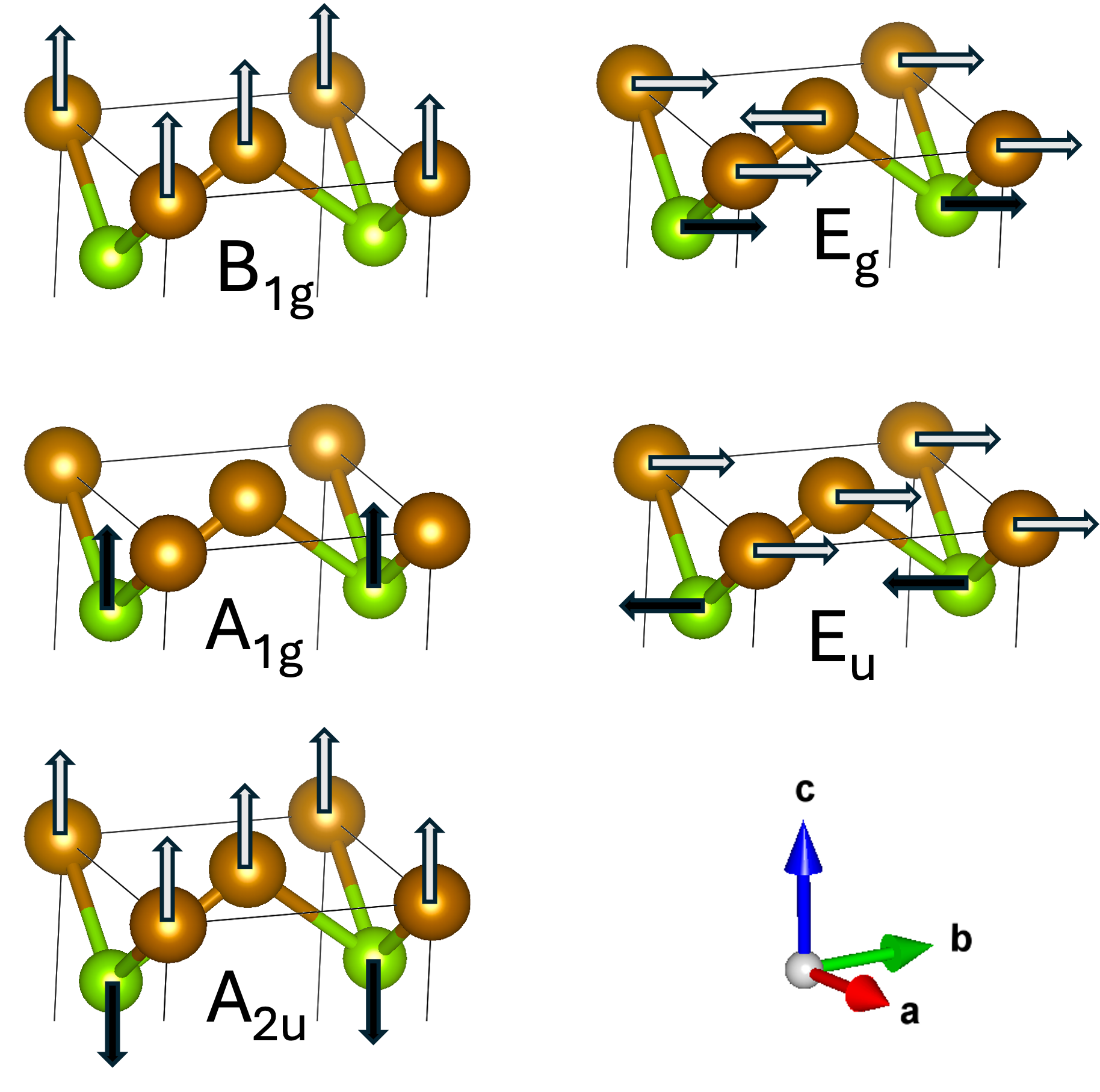}
\caption{Sketch of phonon modes in \ce{FeSe}. The Fe and Se atoms are represented by \enquote{golden} and \enquote{green} sphere.}  
\label{modes}
\end{figure}
Figure~\ref{modes} sketches the phonon modes in \ce{FeSe}. Two fundamental modes are in plane (plane defined by the $a$ and $b$ axes):
\begin{description}
    \item[$E_g$] the Fe atom at the center of the square lattice vibrates in opposite direction to the other Fe atoms and the Se atoms;
    \item[$E_u$] all Fe atoms vibrate in-phase. Se atoms vibrate in the opposite direction.
\end{description}
Three fundamental modes are out-of-plane (along the $c$-axis direction):
\begin{description}
    \item[$B_{1g}$] Fe atoms vibrates in phase;
    \item[$A_{1g}$] Se atoms vibrates in phase; 
    \item[$A_{2u}$] Fe atoms vibrates in phase opposite to the direction of the Se atoms.
\end{description}
%
 The change in the phonon dispersion and DOS with external hydrostatic pressure is shown in Fig.~\ref{press}(b). At higher pressure, acoustic and optical phonon bands move towards higher frequencies---phonon hardening---around both the $\Gamma$ and the $M$ points.
 As can be seen from the total phonon DOS, the shift in phonon frequency is small at lower frequencies (0-20 meV) and gradually becomes larger in the higher frequency region (30-40 meV). Also, with pressure the DOS is reduced at low frequencies and slightly increased at higher pressure. Table~\ref{mode1} lists the energy of the phonon modes with pressure. At ambient pressure in NM phase, the energy for $A_{2u}$ mode (40.0 meV) is in good agreement with experiment, the energy of the pure Se derived mode $A_{1g}$ (16.43 meV) is underestimated, while the energy of the pure Fe-mode $B_{1g}$ (29.57 meV) and of in-plane mode $E_g/E_u$ (35.73 meV) are slightly overestimated. As already observed from Fig.~\ref{press}, external hydrostatic pressure induces hardening of all phonon modes. The largest shift in phonon frequencies is observed when passing from ambient pressure to 2 GPa, while further increase in pressure induces smaller changes. The importance of magnetism on the properties of FeSe has been stressed in a number of studies \cite{PhysRevB.100.064515}. To address the issue, in what follows, we study the dependence of lattice dynamics of long-range magnetic phases of \ce{FeSe} with the external hydrostatic pressure and local magnetic moment. First, we discuss the dynamical stability of the long-range magnetic orders at different pressure and magnetization. Then, the change of the phonon frequencies with magnetization allows us to deduce the spin-phonon coupling and its evolution with pressure and alloying.

\begin{table}[!htbp]
\scriptsize
\tabcolsep=0.04cm
\caption{Frequency (meV) of the phonon modes in NM and SAFM phase and their change with pressure at Fe magnetic moment value 1.9$\mu_B$}
\label{mode1}
\centering
\begin{tabular}{|c|c|c|c|c|c|c|c|c|c|}
\hline
 & \multicolumn{3}{c|}{0 GPa} & \multicolumn{2}{c|}{2 GPa} & \multicolumn{2}{c|}{4 GPa} & \multicolumn{2}{c|}{6 GPa} \\ 
\cline{2-10}
Mode & NM & SAFM & Expt\cite{PhysRevB.96.094531} & NM & SAFM & NM & SAFM & NM & SAFM \\ 
\hline
$A_{1g}$  & 16.43 & 16.74 & 20.5 & 18.04 & 17.61 & 18.65 & 17.85 & 18.72 & 18.06 \\
$B_{1g}$  & 29.57 & 27.84 & 25.6 & 32.59 & 30.91 & 32.90 & 31.11 & 33.51 & 31.54 \\
$A_{2u}$ & 40.0 & 40.17 & 40.0 & 42.39 & 43.68 & 43.13 & 44.42 & 43.81 & 45.08 \\
$E_{g}/E_u$    & 35.73 & 35.42 & 32.0 & 37.04 & 37.40 & 37.64 & 37.58 & 38.32 & 38.18 \\
\hline
\end{tabular}
\end{table}

\subsection{Dynamical stability of long-range magnetic orders}\label{subsec:p}

\begin{figure}
 \centering
 \includegraphics [width=0.30\textwidth]{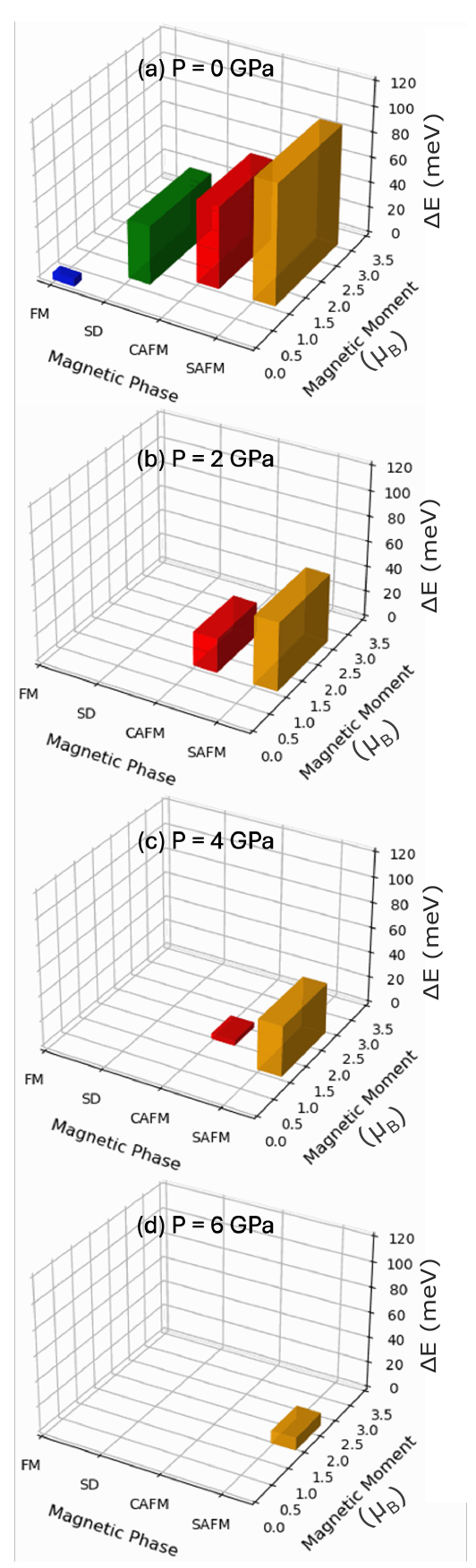}
 \caption{Energetic stability of the FM, SD, CFM and AFM phases (with respect to the nonmagnetic phase) with pressure: (a) 0 GPa, (b) 2 GPa, (c) 4 GPa and (d) 6 GPa and magnetic moment. The stability is quantified in terms of the magnetic formation energy defined as $\Delta E= |E_\text{Magnetic-Phase}-E_\text{Non-Magnetic}|$.}
 \label{stability}
\end{figure}

\begin{figure}
 \centering
 \includegraphics [width=0.48\textwidth]{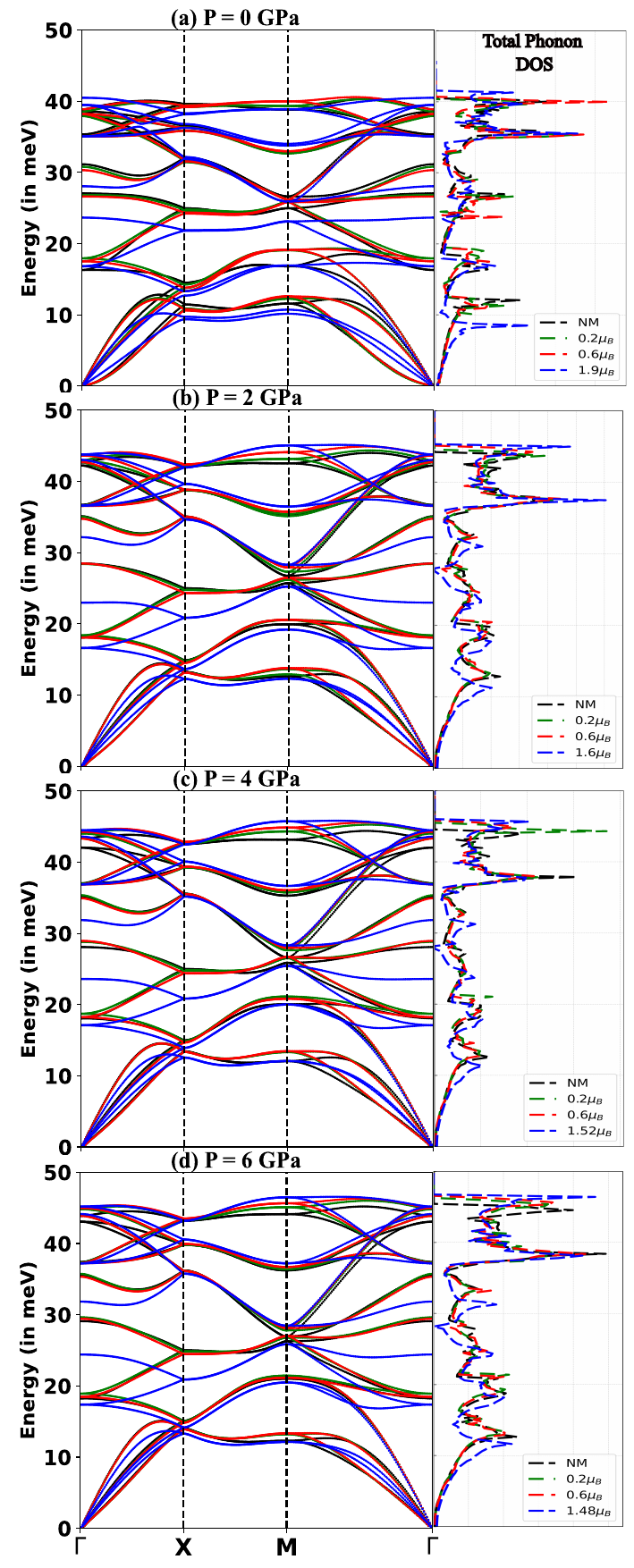}
 \caption{ Phonon dispersion and DOS with different external hydrostatic pressure -- (a) 0 GPa, (b) 2 GPa, (c) 4 GPa and (d) 6 GPa in SAFM phase for three different value of Fe local magnetic moment 0.2$\mu_B$ (\enquote{green}), 0.6$\mu_B$ (\enquote{red}) and higher optimal moment (\enquote{blue}). The nonmagnetic dispersion is shown with \enquote{black} lines.}
 \label{saf}
\end{figure}

\begin{figure}
 \centering
 \includegraphics [width=0.45\textwidth]{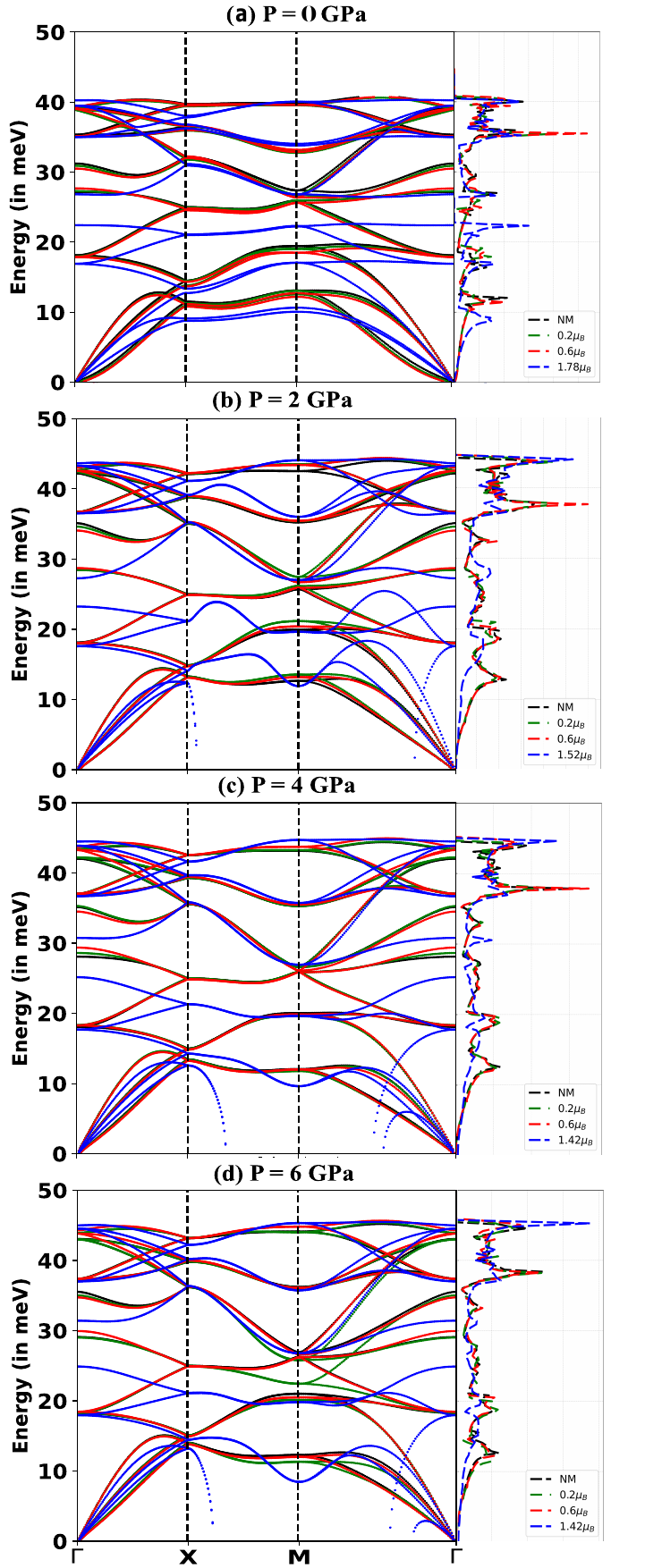}
 \caption{ Phonon dispersion and DOS with different external hydrostatic pressure -- (a) 0 GPa, (b) 2 GPa, (c) 4 GPa and (d) 6 GPa in CAFM phase for three different value of Fe local magnetic moment 0.2$\mu_B$ (\enquote{green}), 0.6$\mu_B$ (\enquote{red}) and higher optimal moment (\enquote{blue}). The nonmagnetic dispersion is shown with \enquote{black} lines.}
 \label{caf}
\end{figure}

\begin{figure}
 \centering
 \includegraphics [width=0.45\textwidth]{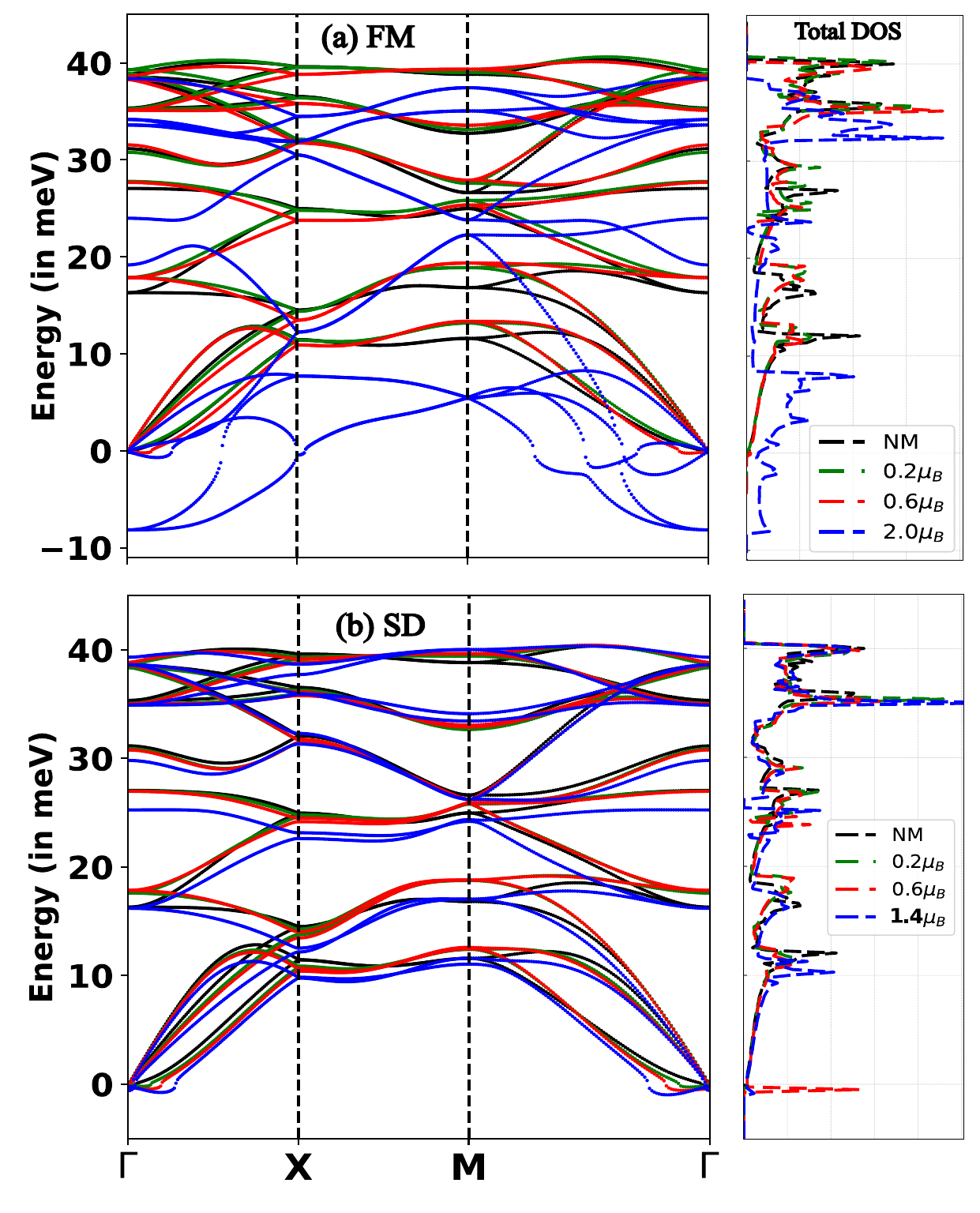}
 \caption{ Phonon dispersion and DOS in (a) FM and (b) SD phase for three different value of Fe local magnetic moment 0.2$\mu_B$ (\enquote{green}), 0.6$\mu_B$ (\enquote{red}) and higher optimal moment (\enquote{blue}) at ambient pressure. The nonmagnetic dispersion is shown with \enquote{black} lines.}
 \label{fm}
\end{figure}

The energetic stability of magnetic phases of \ce{FeSe} has been thoroughly investigated in relation to the ``missing'' long-range magnetic order. In particular, DFT predicts \ce{FeSe} at low local magnetic moment is ferromagnetic, while at higher magnetic moment there is a competition of several AFM phases. Of those, the SAFM and CAFM are energetically stable for increasing pressures (see Fig.~\ref{stability} summarizing results from Ref.~\cite{PhysRevResearch.6.043154}). \emph{To the best of our knowledge, this work presents the first comprehensive investigation of the dynamic stability of magnetic phases in \ce{FeSe}} with pressure. Figures~\ref{saf}, \ref{caf}, \ref{fm}, show the phonon dispersion in the SAFM, CAFM, FM and SD phases for the relevant pressures and magnetic moments. The SAFM phase (Fig.~\ref{saf}) is dynamically stable for all values of pressures and magnetic moments considered.  The CAFM (Fig.~\ref{caf}) is stable at ambient pressure. At higher pressures, the phonon dispersion along $X-M-\Gamma$ softens to negative values. Therefore, above 2 GPa pressure, the CAFM phase is dynamically stable only for low magnetic moments, but not at what is found to be its optimal magnetic moment energetically. 

At ambient pressure, the FM phase [Fig.~\ref{fm}(a)] is dynamically stable only at 0.2$\mu_B$. In fact, the dispersion around $\Gamma$-point reveals phonon softening at M=0.6$\mu_B$ and as we move to high spin state (M=2.0$\mu_B$), frequencies become clearly negative. For the SD phase, the softening around the $\Gamma$-point in the dispersion appears at M=0.2$\mu_B$, which indicates an incipient structural instability of \ce{FeSe} in this phase. At higher magnetization, the SD is dynamically unstable. Therefore, although several possible magnetic phases have been considered to exist in \ce{FeSe} at ambient pressure, only SAFM and CAFM are feasible at ambient pressure and high local magnetic moment. As the pressure increases, it becomes dynamically unfeasible to host CAFM and striped-type spin-density wave appears in the phase diagram, in agreement with experimental observations~\cite{Sun2016}. Note that, for FM and SD phase, the dynamic instability appears for a non-zero momenta that are very close to $\Gamma$ point. This suggests that nanoscopic domains of the FM, SD magnetic phases---as argued in Ref.~\cite{PhysRevResearch.6.043154}---can exist in \ce{FeSe} provided that their size is such that the critical momentum is not admissible due to particle-in-the-box quantization condition.

\subsection{Effect of local magnetic moment on phonon modes}\label{subsec:m}
Overall, in Figures~\ref{saf},\ref{caf},\ref{fm} the dependence of phonon dispersion on magnetic moment hints towards strong magneto-elastic coupling, in agreement with what is observed for example in Ref.~\cite{Gillard2024}.
For the optimal local magnetic moment (1.9$\mu_B$), the phonon frequencies extracted from Fig.~\ref{saf}, are reported in Table.~\ref{mode1}. When compared with the NM phase, the most striking difference is the softening of the $B_{1g}$ mode. 
Similar to the NM phase, there is a hardening of all modes with increasing pressure. This can be associated with the lattice parameter $c/a$ ratio and $z_{\ce{Se}}$. As the pressure increases from ambient to 6 GPa, $c/a$ changes from 1.49 to 1.43. In contrast to the NM phase, the increase in frequency when increasing the pressure is more pronounced for the $A_{2u}$ mode. Also, frequency of the $A_{1g}$ mode increases when going from low to high pressure. The $A_{1g}$ mode is primarily associated with $z_{\ce{Se}}$ which increases with both long range magnetic order as well as local magnetic moment.

\begin{table}[!htbp]
\scriptsize
\tabcolsep=0.15cm
\caption{Frequency (meV) of the phonon modes in CAFM, SD and FM phase at optimum magnetic moment value for a particular phase.}
\label{fm1}
\centering
\begin{tabular}{|c|c|c|c|c|}
\hline
Phonon Mode & NM & CAFM  & SD & FM  \\ 
\hline
$A_{1g}$ & 16.43& 17.00  & 16.74  & 18.52 \\
$B_{1g}$ & 29.57 & 26.67  & 28.45 & 29.42 \\
$A_{2u}$ & 40.0 & 38.89  & 39.89 & 39.95 \\
$E_{g}/E_u$ & 35.73  & 35.02  & 34.89 & 35.48 \\
\hline  
\end{tabular}
\end{table}
Table~\ref{fm1} lists the energy of the phonon modes in the CAFM, FM and SD phase at ambient pressure and M=1.7$\mu_B$, 0.2$\mu_B$ and 1.4$\mu_B$ respectively. With respect to the NM phase, the largest differences are observed for the $A_{1g}$ and $B_{1g}$ modes. In the FM phase, the $A_{1g}$ is significantly higher in frequency (18.52 meV). In the CAFM, the frequency of the $B_{1g}$ mode (26.67 meV) softens even compared to the SAFM phase. Again, these differences can be rationalized from changes in $z_{\ce{Se}}$. The value of $z_{\ce{Se}}$ in CAFM phase is lower than that in SAFM~\cite{PhysRevResearch.6.043154}. This reduction in $z_{\ce{Se}}$ correlates with a further softening of both the $A_{1g}$ and $B_{1g}$ modes in the CAFM phase. Note that $z_{\ce{Se}}$ in CAFM at ambient pressure calculated within DFT is closer to the experimentally measured value and so is the frequency of the $B_{1g}$ mode (Tab.~\ref{mode1}).

\subsection{Effect of S/Te substitution on phonon modes}\label{subsec:s}
\begin{figure}
 \centering
 \includegraphics [width=0.4\textwidth]{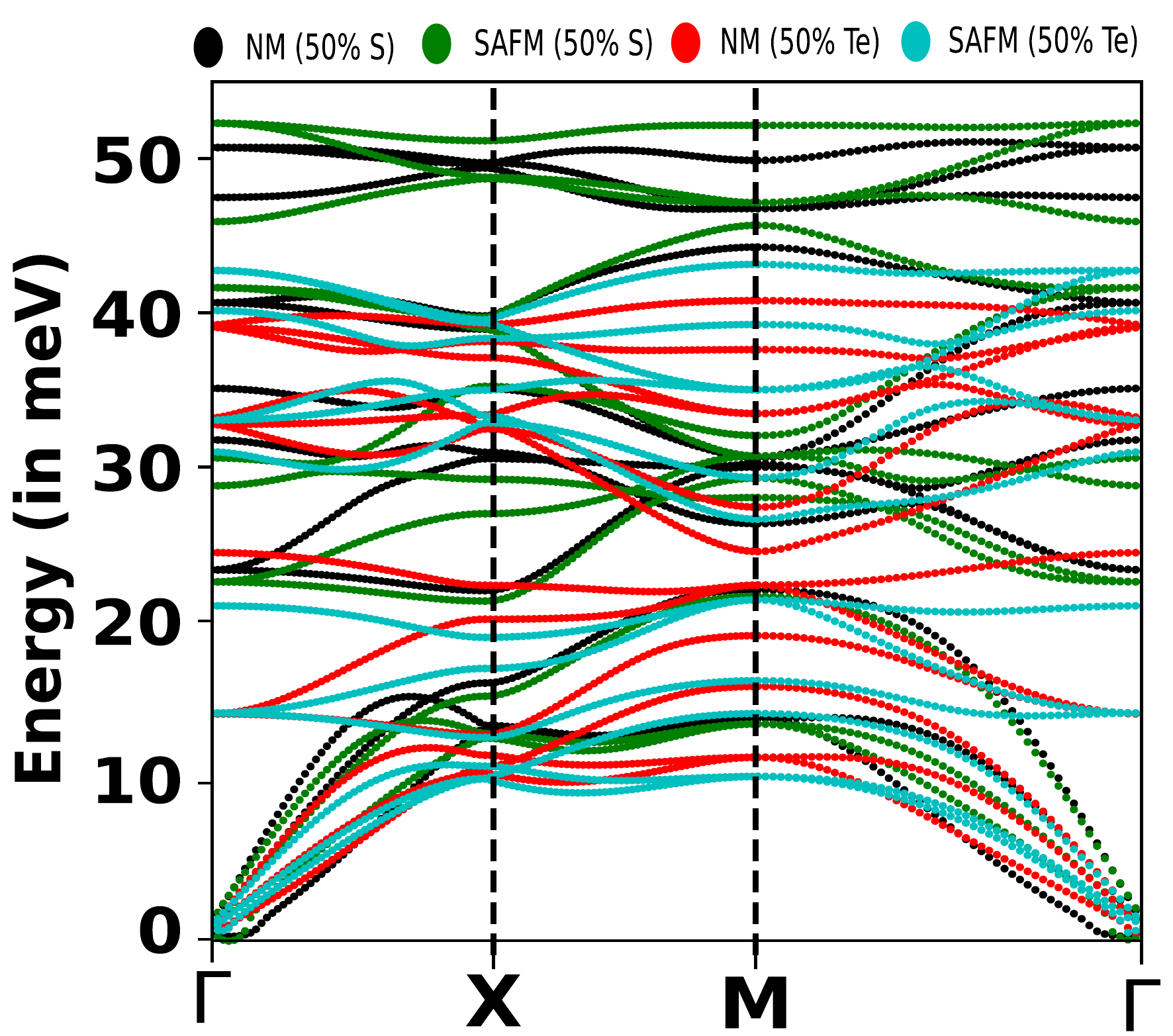}
 \caption{ Phonon band structure at 50\% S/Te substitution at Se site for NM and SAFM phase.} 
 \label{ste}
\end{figure}
The phonon dispersion with 50\% of both S and Te substitution at Se site is presented in Fig.\ref{ste} for the NM and SAFM phases. The two kinds of substitution have a markedly different effect on the phonon band dispersion. Due to Te substitution, all the phonon modes shift towards lower frequency owing to larger atomic mass of Te. Conversely, the lower atomic mass of S (in comparison to that of Se) shifts the phonon modes to higher frequency relative to that of pristine FeSe. When comparing the NM and SAFM phases, the presence of magnetic order affects dispersion with both kind of substitution visibly. The frequency modes (particularly around M-point) approximately below 25 meV shifts to lower frequency. In contrast, frequency modes above 25 meV move to higher frequency. The $A_{1g}$ mode in the NM phase has a frequency of 23.31 (13.92) meV for 50\% S(Te) substitution compared to NM-FeSe (16.43 meV). In SAFM phase, the frequency shifts to 22.48 (14.07) meV for (S/Te) alloyed FeSe compared to 16.74 meV in pristine case. The $B_{1g}$ mode shifts towards higher frequency---35.08 (32.64) meV for S(Te) substitution compared to 29.57 meV in pristine NM case. Like pristine FeSe, SAFM pushes the $B_{1g}$ mode to a lower frequency -- 30.45(30.97) meV for S(Te) substitution.

\subsection{Spin-phonon coupling}\label{sec:spc}

\begin{figure}
 \centering
 \includegraphics [width=0.45\textwidth]{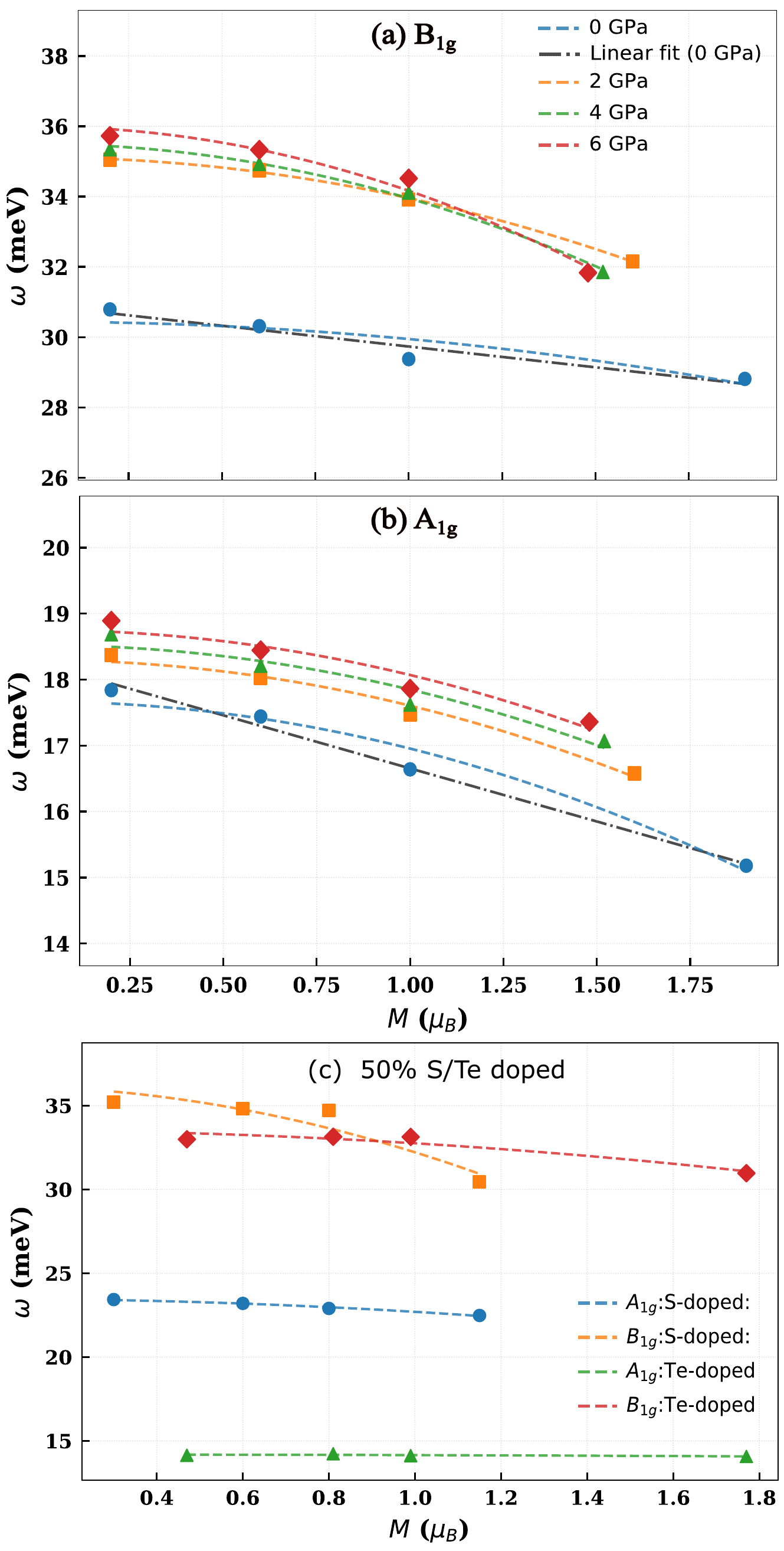}
 \caption{Linear regression using the model $\omega = a_0+a_2M^2$ for the (a) $B_{1g}$ and (b) $A_{1g}$ mode at 0 GPa (blue  circle),2 GPa (orange square), 4 GPa (green triangle) and 6 GPa (red diamond). The fit to the linear model $\omega = a_0+a_2M$ at $P=0$~GPa is shown with black dashed line; (c) the same with 50\% S/Te substitution. }
 \label{fit}
\end{figure}

\begin{table}[!htbp]
\scriptsize
\tabcolsep=0.10cm
\caption{Fitting coefficients from the linear regression of the quadratic ($\omega = a_0+a_2M^2$) and the linear ($\omega = a_0+a_1M$) model for the dependence of the phonon frequency of the $A_{1g}$ and $B_{1g}$ modes on the magnetization $M$ at different pressures ($P$). The $R^2$ value is also reported. For the $a_0$, in parenthesis is given the difference with the corresponding phonon frequency in the NM phase.}
\label{fita}
\centering
\begin{tabular}{|c|c|c|c|c|c|c|}
\hline
 & \multicolumn{3}{c|}{$B_{1g}$} & \multicolumn{3}{c|}{$A_{1g}$} \\ 
\cline{2-7}
$P$ (GPa) & $a_0$ & $a_{2/1}$ & $R^2$ & $a_0$ & $a_{2/1}$ & $R^2$ \\ 
\hline
0  & 30.44 (0.87)& -0.49 & 0.80 & 17.67 (1.24) & -0.71 & 0.96 \\

0 (linear)  & 30.92 (1.35) & -1.18 & 0.93 & 18.26 (1.83) & -1.61 & 0.99 \\
\hline
2  & 35.11 (2.52) & -1.16 & 0.99 & 18.30 (0.26)& -0.69 & 0.98 \\
2 (linear)  & 35.77 (3.18) & -2.12 & 0.94 & 18.72 (0.68)& -1.30 & 0.98 \\
\hline
4  & 35.50 (2.60) & -1.55 & 0.99 & 18.52 (-0.04)& -0.67 & 0.93 \\
4 (linear)  & 36.26 (3.36) & -2.64 & 0.91 & 18.93 (0.37) & -1.24 & 0.99 \\
\hline
6  & 35.99 (2.48)& -1.83 & 0.98 & 18.75 (0.03) & -0.68 & 0.94 \\
6 (linear)  & 36.81 (3.30) & -3.0 & 0.87 & 19.14 (0.42)& -1.22 & 0.99 \\
\hline
50\% S  & 36.20 (1.12)& -3.98 & 0.886 & 23.47 (0.16) & -0.77 & 0.99 \\
50\% S (linear)  & 36.69 (1.61)& -5.46 & 0.756 & 23.82 (0.51) & -1.14 & 0.986 \\
\hline
50\% Te   & 33.53 (0.89) & -0.78 & 0.91 & 14.18 (0.26)& -0.04 & 0.48 \\
50\% Te (linear)  & 34.30 (1.66)& -1.72 & 0.801 & 14.22 (0.30) & -0.08 & 0.411 \\
\hline
\end{tabular}
\end{table}

The magnetic moment-dependent renormalization of phonon dispersion in \ce{FeSe} resembles that of FeSi, where strong spin-phonon coupling is evident ~\cite{Krannich2015}. This similarity suggests strong spin-phonon coupling in FeSe, which might play a key role in the determination of the superconducting properties of this compound. Renormalized phonon frequency due to presence of long-range magnetic order can be computed from the real-part of self energy~\cite{PhysRevB.99.184442}. In Ref.~\cite{10.1063/1.342186}, it was shown that (in the lowest order) this can be expressed for a system with long range magnetic order at T=0~K as:
\begin{equation}
\omega_p = \omega_0 -g S_z^2,
\end{equation}
where $\omega_0$ is the frequency in absence of spin-phonon coupling, $g$ quantifies the spin-phonon coupling and the negative sign comes due to antiferromagnetic ordering. In Fig.~\ref{fit}, we consider the polynomial fitting of $\omega$ versus M with the model $\omega=a_0+a_2M^2$ for the pure Fe $B_{1g}$ and pure Se $A_{1g}$ phonon modes of the SAFM phase. We consider this phase as it remains energetically and dynamically stable throughout the pressure and magnetic moment range. The calculated fitting parameters are listed in Tab.~\ref{fita} together with the $R^2\rightarrow1$ value indicating how well the model fits the results. For comparison, the parameters and $R^2$ are also reported for a linear model. Since, $a_0$ provides the frequency at $M=0$~$\mu_B$, we compare its value with that of the phonon frequency in the NM mode (Tab.~\ref{mode1}). The values in brackets with $a_0$ show the difference between the two. As $a_2$ is extracted from energy variation of phonon modes with magnetic moment, it contains information about coupling of spin-orbital with phonon.

In Fig.~\ref{fit}(a,b), we see that pressure affects the two modes differently. The effect of pressure is very prominent on $B_{1g}$ mode. For the $B_{1g}$ mode, at ambient pressure, a linear fit works better than a quadratic one by comparing the value of $R^2$ (see $P=0$~GPa in Tab.~\ref{fita}). As the pressure increases, the quadratic model fits the results better than the linear one. This hints towards a change in the nature of spin-phonon coupling in \ce{FeSe} at higher pressure. At 2 GPa pressure, the dependence of the $B_{1g}$ frequency on $M$ becomes stronger compared to that at ambient pressure, as reflected by the increase in the quadratic coefficient  $a_2$ from 0.49 to 1.16. At higher pressures (4~GPa, 6~GPa), $a_2$ increases significantly indicating an increase in the spin-phonon coupling. The $A_{1g}$ mode exhibits a stronger magnetic moment dependence of phonon frequency compared to $B_{1g}$  at ambient pressure. This suggests a prominent spin-phonon coupling mediated by the \ce{Se}--\ce{Fe}--\ce{Se} ($z_{\ce{Se}}$) Kramers-Anderson super-exchange pathway, similar to the two-dimensional antiferromagnet \ce{MnPSe_3}~\cite{Gillard2024}. 

As pressure increases, the curves fitting the phonon frequency of the $A_{1g}$ mode on the magnetic-moment are very similar but for a shift, implying reduced spin-phonon coupling variation, as can be seen from the minimal changes in the coefficient $a_2$ with pressure. Notice that unlike the $B_{1g}$ mode, the difference in $R^2$ between the linear and quadratic fit for the $A_{1g}$ mode is less prominent. On the other hand, the value of $a_0$ (the phonon frequency at zero magnetic-moment) is still closer to $\omega_0$ (the corresponding frequency in the nonmagnetic phase) for the quadratic fit throughout the pressure range. 

The fit for 50\% S/Te substitution in FeSe for the two phonon modes is presented in Fig.~\ref{fit}(c). For 50\% S-substitution in FeSe, the value of $a_2(g)$ obtained from the quadratic fit increases remarkably for $B_{1g}$ mode due to reduced anion height ($z_S$). On the other hand, for Te the larger anion-distance, $z_{Te}$, reduces the value of $a_2$ ($g$).

\subsection{Discussions}\label{sec:spcdis}
Our results show that the nature of spin-phonon coupling evolves substantially under variations in pressure and alloying. Based on these results, we can infer the potential role of spin-phonon coupling in shaping the phase diagram of FeSe. At ambient pressure, when \ce{FeSe} hosts nematic phase and lacks long-range order, spin-phonon coupling of both $A_{1g}$ and $B_{1g}$ modes is low and of comparable magnitude. In contrast, at higher pressures, when the striped long-range order develops into the system, spin-phonon coupling of the $B_{1g}$ becomes stronger (about 1.5 to 2.5 times). Above 6 GPa, spin-phonon coupling is further enhanced and this may be the reason behind the suppression of striped phase. Therefore, It can be inferred that a moderate to strong spin-phonon coupling is necessary to host long range order. At 50\% S-substitution case, an exceptionally high spin-phonon coupling for the $B_{1g}$ mode may suppress both nematicity as well as stripe type order in \ce{FeSe}. On the contrary, low \enquote{$g$} for 50\% Te-substitution case signals towards weak stripe type fluctuations which is confirmed by experiments \cite{cite-key5}. Low \enquote{$g$} also implies persistence of nematic fluctuations even at high Te-concentration in \ce{FeSe} as reflected in the phase diagram. Together, these results demonstrate that both $z_{\ce{Se}}$ and the lattice parameters play a key role in determining magneto-elastic coupling which in turn influences the superconducting property of \ce{FeSe}, similar to what is observed in iron pnictide compounds \cite{PhysRevB.86.134507}.

Regarding the linear dependence of the $B_{1g}$ mode frequency with $M$ at ambient pressure, this can be rationalised by the field-theory model of two-dimensional spin and orbital fluctuations we put forward in Ref.~\cite{PhysRevResearch.6.043154,s5cy-t1sf}.  Indeed, the linear dependency may be a manifestation of the degeneracy of two bosonic modes: the lattice phonon and orbital fluctuation. As the system approaches dispersion crossing, interconversion is allowed (by momentum-energy conservation) and bosons hybridize into a mixed particle~\cite{PhysRevB.99.184442}. As an evidence to support this hypothesis, in Ref.~\cite{PhysRevResearch.6.043154} we found that at ambient pressure and low magnetization, the energy difference between the lowest phase (FM) and the CAFM/SAFM phases is $\approx 30$~meV. At higher pressures this energy difference decreases. According to the model, the energy difference between phases corresponds roughly to the energy of the orbital fluctuation, which would be indeed close in energy and mix with the lattice phonon at ambient pressure. At higher pressure instead, the energy of the two bosonic modes is substantially different with no mixing occurring. The absence of mixing at higher pressure, would contribute to stabilize the orbital order, in agreement with Ref.~\cite{PhysRevResearch.6.043154}.

\section{Conclusions}\label{sec:con}
With the exploration of magnetism-dependent phononic band structure, we gained insight into the interplay between magnetism and lattice dynamics in FeSe instrumental to rationalize the FeSe phase diagram. Phonon calculations show that most energetically competitive magnetic phases are dynamically unstable even at ambient pressure, except for the stripe antiferromagnetic phase. This behavior is consistent with the experimentally observed evolution from  mixed stripe-N\'eel-type fluctuations at ambient pressure to stripe-type order at higher pressures ~\cite{Sun2016}. Our analysis also reveals a pronounced pressure and alloying tunable spin–phonon coupling. Changes in structural parameters, particularly the Se height, modify magnetic interactions and thereby alter both the nature and magnitude of spin-phonon coupling for the $A_{1g}$ and $B_{1g}$ phonon modes. Finally, with our results we can interpret the evolution of magnetic phases with pressure and alloying in terms of the coupling between the spin and phonon degrees of freedom. Specifically, long-range magnetic order appears for moderate-to-strong spin–phonon coupling, while it is suppressed for either too low or too high coupling.   

\begin{acknowledgments}
The authors are grateful for the use of the computing resources from the Northern Ireland High Performance Computing (NI-HPC) service funded by EPSRC (EP/T022175); and the ARCHER2 UK National Supercomputing Service (https://www.archer2.ac.uk)~\cite{beckett2024archer2} through the UKCP consortium funded by EPSRC (EP/X035891/1). PC acknowledges financial support of NCN-PL through grant No. 2021/43/B/ST8/03207. The authors thank Nigel Hussey and M. A. Fadla for useful discussions. This work was supported by the Engineering and Physical Sciences Research Council (EPSRC), under grant EP/V029908/1. 

\end{acknowledgments}

\bibliography{FeSe}
\end{document}